\documentclass{article}

\usepackage{amsmath}
\usepackage{amssymb}
\usepackage{graphics}
\usepackage{url}
\usepackage{algpseudocode}

\newcommand{\Rule}{\mathcal{R}}
\newcommand{\fail}{\mathsf{fail}}
\newcommand{\nl}{\mathord{!}}

\newcommand{\Goto}[1]{\textbf{goto}~#1}
\algblockdefx[SWITCH]{Switch}{EndSwitch}
	[1]{\textbf{switch}~#1}
	[0]{\textbf{end switch}}

\algblockdefx[CASE]{Case}{EndCase}
	[1]{\textbf{case}~#1:}

\algblockdefx[STATICCASE]{StaticCase}{EndStaticCase}
	[1]{#1 $\implies$}
	[1]{#1}

\newcommand{\etal}{\textit{et~al.}\@}

\title{Parsing Expression GLL}
\author{
    Moss, Aaron\thanks{This work was funded by a Shiley grant at the University of Portland.}\\
    \texttt{mossa@up.edu}
    \and
    Harrington, Brynn\\
    \texttt{harringt23@up.edu}
    \and
    Hoppe, Emily\\
    \texttt{hoppe23@up.edu}
}

\begin{document}
\maketitle

\begin{abstract}
This paper presents an extension of the GLL parsing algorithm for context-free grammars which also supports parsing expression grammars with ordered choice and lookahead. 
The new PEGLL algorithm retains support for unordered choice, and thus parses a common superset of context-free grammars and parsing expression grammars. 
As part of this work, the authors have modified an existing GLL parser-generator to support parsing expression grammars, adding operators for common parsing expressions and modifying the lexer algorithm to better support ordered choice. 
\end{abstract}

\section{Introduction}
The inherently unambiguous nature of parsing expression grammars (PEGs) makes them an attractive choice for modelling structured text such as programming languages and computing data formats, but in practice the superior performance of parsers based on context-free grammars (CFGs) has led to CFGs being more-widely used, despite the difficulty of disambiguating them. 
This work is an initial effort toward a unified framework that provides the advantages of both grammar formalisms: it is an algorithm adopted from an efficient, general-purpose CFG parser that supports PEG semantics without discarding support for the unordered choice operator of CFGs in cases where that ambiguity may be desirable. 

More specifically, this paper presents a modification of the GLL parser-generator of Scott \& Johnstone\cite{SJ10,SJ16}. 
The key contribution is the \emph{FailCRF} data structure, which adds a failure path to Scott \& Johnstone's call-return forest; the addition of a failure path allows the lookahead and ordered choice operations of the PEG formalism to be supported.
The authors have extended Ackerman's GoGLL\cite{Ack19} parser-generator to implement this new algorithm, adding syntactic sugar for common PEG operators and modifying the lexer algorithm to allow the PEG parser to override the usual maximal-munch rule. 

\section{Parsing Expression Grammars}
The primary difference between parsing expression grammars and the more familiar context-free grammars is \emph{ordered choice}: PEGs, as a formalism of recursive-descent parsing introduced by Ford \cite{For02}, do not try subsequent alternatives of an alternation if an earlier alternative matches. 
The other significant difference between the PEG and CFG formalisms are the PEG \emph{lookahead} expressions, $\nl\alpha$ and $\&\alpha$, which match only if the subexpression $\alpha$ does not (resp. does) match, but consume no input regardless. 
These lookahead operators provide the infinite lookahead of the PEG formalism.
The other fundamental PEG operators act much like their CFG equivalents, and are described in Fig.~\ref{expr-fig} as functions over an input string $s$ drawn from some alphabet $\Sigma$ producing either a (matching) suffix of $s$ or the special value $\fail \not\in \Sigma^*$.
In summary, the \emph{string literal} $u$ matches and consumes the string $u$, the \emph{empty expression} $\varepsilon$ always matches without consuming anything, while the \emph{failure expression} $\varnothing$ never matches. 
A \emph{nonterminal} $A$ is replaced by the parsing expression $\Rule(A)$ it corresponds to.
The \emph{sequence} expression $\alpha\beta$ matches $\alpha$ followed by $\beta$, while the \emph{ordered choice} expression $\alpha/\beta$ only tries $\beta$ if $\alpha$ does not match. 
To differentiate CFG \emph{unordered choice}, it is represented in this paper as $\alpha|\beta$.

\begin{figure}
	\centering
	\begin{equation*}
	\begin{aligned}[c]
	u(s)            & = \begin{cases} v     & s = uv \\
	                                  \fail & \text{otherwise} \end{cases} \\
	\varepsilon(s)  & = s \\
	\varnothing(s)  & = \fail \\
	\nl\alpha(s)    & = \begin{cases} s     & \alpha(s) = \fail \\
	                                  \fail & \text{otherwise} \end{cases} \\
    \&\alpha(s)     & = \begin{cases} s     & \alpha(s) \neq \fail \\
                                      \fail & \text{otherwise} \end{cases}
	\end{aligned}
	~~~~
	\begin{aligned}[c]
	A(s)            & = (\Rule(A))(s) \\
	\alpha\beta(s)  & = \begin{cases} \beta(\alpha(s)) & \alpha(s) \neq \fail \\
	                                  \fail            & \text{otherwise} \end{cases} \\
	\alpha/\beta(s) & = \begin{cases} \alpha(s)  & \alpha(s) \neq \fail \\
	                                  \beta(s)   & \text{otherwise} \end{cases}
	\end{aligned}
	\end{equation*}
	\caption[Expression definitions]{Formal definitions of parsing expressions} \label{expr-fig}
\end{figure}

\section{GLL Parsing}
\emph{Generalized LL} (GLL) parsing, introduced by Scott \& Johnstone\cite{SJ10,SJvB19}, extends the power of LL parsing to all CFGs through use of a \emph{call-return forest} (CRF) to represent the recursive-descent call stack of the LL parsing algorithm. 
For efficiency, the CRF is implemented using the \emph{graph-structured stack} (GSS) data structure introduced by Tomita\cite{Tom85} for the GLR parsing algorithm. 
The gist of the GLL approach is that each CRF node represents a function call (equivalently, nonterminal invocation) in a recursive-descent parse, and includes an input position, a nonterminal to match, and a grammar slot to return to on completion. 
The graph structure of this stack comes from a dynamic de-duplication of CRF nodes which share a nonterminal and input position, changing a stack data structure into a directed acyclic graph (DAG). 
The GLL algorithm keeps a queue of CRF nodes which are pending parsing, and handles the nondeterminism of unordered choice by enqueuing a CRF node for each choice.

Scott \etal\cite{SJvB19} introduced \emph{binary subtree representation} (BSR) sets as an output format to represent nonterminal matches in GLL. 
The essential insight is that, while the traditional \emph{shared packed parse forest} (SPPF)\cite{Tom85} data structure representing possible parse trees requires significant complication in the parser algorithm to properly store and update edges between parse tree nodes, those edges can be efficiently reconstructed from an indexed set of edgeless parse-tree nodes (the BSR set) with minimal added information. 

A BSR element is a 4-tuple containing a \emph{grammar slot} $X ::= \alpha\theta\cdot\beta$, and three input indices $i$, $j$, and $k$, $i \leq j \leq k$. 
The BSR element represents a successful match of the nonterminal $X$ up to the end of $\theta$, the single terminal or nonterminal immediately before the dot of the grammar slot; $i$ is the input index where $X$ began to match, $j$ is the index where $\theta$ began to match, and $k$ is the index where $\theta$ finished matching. 
Note that if $\beta = \varepsilon$, the BSR node represents a complete match of $X$.
Parse trees can be straightforwardly reconstructed from BSR sets: a predecessor of a BSR element $(X ::= \alpha\theta\cdot\beta, i, j, k)$ is any element $(X ::= \alpha\cdot\theta\beta, i, \ell, j)$, while its child where $\theta$ is some nonterminal $A$ is any element $(A ::= \delta\cdot, j, m, k)$. Successor and parent elements can be defined analogously.

\section{Parsing Expression GLL}
The \emph{Parsing Expression GLL} (PEGLL) algorithm introduced in this paper uses similar data structures and abstractions as GLL for CFGs. 
The main loop is outlined in Figure~\ref{main-loop-algo}; it first initializes an empty queue $R$ of slot descriptors to parse, an empty cache $U$ of previously seen slot descriptors, and an empty set $T$ of BSR elements to report.
It then queues the start rule of the grammar at input position 0 for parsing, parses each descriptor, and completes by returning whether or not the start rule matched at position 0. 
Note that (unlike CFGs), PEGs match prefixes of their input, so the start rule may only consume the input up to some index $k$; if this behavior is not desired, a match can be returned only if $k$ is the length of the input string.

\begin{figure}
\caption{PEGLL main loop} \label{main-loop-algo}
\begin{algorithmic}
\State $R \gets \emptyset$, $U \gets \emptyset$, $T \gets \emptyset$
\State \Call{addNt}{$S$, 0}
\While{$R \neq \emptyset$} 
    \State $(L, c_U, c_I) \gets R$.\Call{remove}{}
    \State $t \gets$ \Call{tokens}{$c_I$} $t' \gets t$
    \Loop
        \Switch{$L$}
            \State $\langle$ generate code for each rule $R$ $\rangle$
        \EndSwitch
    \State nextSlot: \EndLoop
\State nextDesc: \EndWhile
\If{$\exists \alpha, j, k, (S ::= \alpha \cdot, 0, j, k) \in T$}
    \State \Return match at maximal such $k$
\Else
    \State \Return $\fail$;
\EndIf
\end{algorithmic}
\end{figure}

The primary thing the loop in Figure~\ref{main-loop-algo} does is dispatch the current descriptor to the code which executes its parse. 
The code for each nonterminal may be generated according to the patterns in Figures~\ref{nt-pattern} and~\ref{empty-nt-pattern}. 
This parser-generator assumes any ordered choice expressions are at the very top level of a nonterminal (parenthesized subexpressions are added as syntactic sugar, see Section~\ref{syntax-sec}). 
A label is then generated for each alternate, failing over to the next alternate if it does not match, with a synthesized failure alternate at the end of each alternation.

\begin{figure}
\caption[Non-terminal code pattern]{Pattern for a nonterminal $X ::= \tau_1 / \ldots / \tau_p, p \geq 1$} \label{nt-pattern}
\begin{algorithmic}
\Case{$X ::= \cdot \tau_1$}
	\State $\langle$ generate code for $X ::= \cdot \tau_1$ with failure path $X ::= \cdot \tau_2$ $\rangle$
	\State \Call{rtn}{$X$, $c_U$, $c_I$}; \Goto{nextDesc}
\EndCase
\State $\vdots$
\State ~
\Case{$X ::= \cdot \tau_p$}
	\State $\langle$ generate code for $X ::= \cdot \tau_p$ with failure path $X ::= \cdot \varnothing$ $\rangle$
	\State \Call{rtn}{$X$, $c_U$, $c_I$} \Goto{nextDesc}
\EndCase
\Case{$X ::= \cdot \varnothing$}
	\State \Call{rtn}{$X$, $c_U$, $\fail$} \Goto{nextDesc}
\EndCase
\end{algorithmic}
\end{figure}

\begin{figure}
\caption[Empty nonterminal code pattern]{Pattern for a nonterminal $X ::= \varepsilon$} \label{empty-nt-pattern}
\begin{algorithmic}
\Case{$X ::= \cdot \varepsilon$}
	\State $T \gets T \cup {(X ::= \varepsilon \cdot, c_I, c_I, c_I)}$
	\State \Call{rtn}{$X$, $c_U$, $c_I$} \Goto{nextDesc}
\EndCase
\end{algorithmic}
\end{figure}

The \textsc{rtn}$(X, c_U, c_I)$ function in PEGLL code is detailed below in Figure~\ref{call-rtn-code}, but its essential purpose is to modify the ``call stack'' in the CRF graph consistently with a recursive call to the nonterminal $X$ at position $c_U$ returning at position $c_I$.

To parse each of the sequence expressions $\tau_i$ inside a nonterminal alternative, PEGLL repeatedly executes the appropriate code for each expression in the sequence, moving to the failure path if that expression does not work. 
The code patterns for the sequence expression are in Figure~\ref{seq-expr-pattern}, while the code patterns for each atomic expression are in Figure~\ref{atom-expr-pattern}. 
Terminal matches advance the input index $c_I$ to the right extent $r$ of the token and find the new tokens $t$ at that position, while mid-sequence errors reset $c_I$ and $t$ to their initial values in the descriptor, $c_U$ and $t'$.
Note also that nonterminal calls preserve the failure path from their caller, and in particular that negative-lookahead expressions swap the success and failure results of the call.

\begin{figure}
\caption[Sequence expression code pattern]{Pattern for a sequence $X ::= \cdot x_1 \ldots x_d$ with failure path $L_f$} \label{seq-expr-pattern}
\begin{algorithmic}
\State $r \gets$ \Call{testSelect}{$X ::= \cdot x_1 \ldots x_d$, $t$}
\If{\Call{testSelect}{} failed}
	\State report error
	\State $(L, c_I, t) \gets (L_f, c_U, t')$ \Goto{nextSlot}
\EndIf
\State $\langle$ generate code for $X ::= x_1 \cdot x_2 \ldots x_d$ with failure path $L_f$ $\rangle$
\State $\vdots$
\State $\langle$ repeat for remaining atoms $x_2 \ldots x_d$ in sequence $\rangle$
\end{algorithmic}
\end{figure}

\begin{figure}
\caption[Atomic expression code patterns]{Pattern for an atomic expression $X ::= \alpha \varphi \cdot \beta$ with failure path $L_f$} \label{atom-expr-pattern}
\begin{algorithmic}
\StaticCase{$\varphi = a$}
	\State $R \gets R \cup (X ::= \alpha a \cdot \beta, c_U, c_I, r)$
	\State $c_I \gets r$
	\State $t \gets$ \Call{tokens}{$c_I$}
\EndStaticCase{}
\StaticCase{$\varphi = Y$}
	\State \Call{call}{$X ::= \alpha ~Y \cdot \beta$, $L_f$, $Y$, $c_U$, $c_I$} \Goto{nextDesc}
\EndStaticCase{\textbf{case} $X ::= \alpha ~Y \cdot \beta$:}
\State ~
\StaticCase{$\varphi = \& Y$}
	\State \Call{call}{$X ::= \alpha ~\& Y \cdot \beta$, $L_f$, $Y$, $c_U$, $c_I$} \Goto{nextDesc}
\EndStaticCase{\textbf{case} $X ::= \alpha ~\& Y \cdot \beta$:}
\State ~
\StaticCase{$\varphi = \nl Y$}
	\State \Call{call}{$L_f, X ::= \alpha ~\nl Y \cdot \beta$, $Y$, $c_U$, $c_I$} \Goto{nextDesc}
\EndStaticCase{\textbf{case} $X ::= \alpha ~\nl Y \cdot \beta$:}
\end{algorithmic}
\end{figure}

Together, the helper functions \textsc{call} and \textsc{rtn} simulate the call stack of a recursive-descent PEG parser; full code is in Figure~\ref{call-rtn-code}. 
The significant difference between the \emph{FailCRF} of PEGLL and the \emph{CRF} of Scott~\etal \cite{SJvB19} is that edges in the \emph{FailCRF} are labelled as either ``match'' edges or ``fail'' edges, representing successful and unsuccessful return paths, respectively. 
The motivation for this modification is handling negative lookahead expressions --- $\nl X$ matches only if $X$ does not, and thus the CRF needs to encode the failure of $X$ as the trigger of a move from slot $Y ::= \alpha \cdot \nl X \beta$ to slot $Y ::= \alpha \nl X \cdot \beta$. 
Fail edges are also used to encode the ordered choice rule; while the GLL algorithm attempts to parse all alternates of a nonterminal concurrently, PEGLL attempts one at a time, following its fail edge to the next alternate if that one fails. 

In more detail, the \textsc{call}$(L_m, L_f, X, i, j)$ function enqueues parsing of nonterminal $X$ at position $j$, returning to slot $L_m$ on match or $L_f$ on failure, where $L_m$ or $L_f$ began parsing at position $i$. 
The \textsc{rtn}$(X, j, h)$ function reports the result of parsing nonterminal $X$ begining at position $j$, where $h$ is either the last consumed position or the special value $\fail$ indicating that the parse did not succeed.
For convenience, this presentation assumes the call-return forest (CRF) is stored in a global cache. 
The ``popped cache'' of previously-parsed results is included to allow updating of previous parse results when they are used in a new context and also assumed to be global.

\begin{figure}
\caption{Code for \textsc{call} and \textsc{rtn} functions} \label{call-rtn-code}
\begin{algorithmic}
\Function{call}{slot $L_m$, slot $L_f$, nonterminal $X$, int $i$, int $j$}
	\State $u_m \gets$ CRF node $(L_m, i)$, created if not in cache
	\State $u_f \gets$ CRF node $(L_f, i)$, created if not in cache
	\State $v \gets$ CRF node $(X, j)$, created if not in cache
	\If{$v$ was not previously in cache}
		\State add a match edge from $v$ to $u_m$ and a fail edge from $v$ to $u_f$
		\State \Call{addNt}{$X$, $j$}
	\Else
		\If{there is not a match edge from $v$ to $u_m$}
			\State add a match edge from $v$ to $u_m$
			\ForAll{$(X, j, h), h \neq \fail$ in popped cache}
				\State \Call{addMatch}{$L_m$, $i$, $j$, $h$}
			\EndFor
		\EndIf

		\If{there is not a fail edge from $v$ to $u_f$}
			\State add a fail edge from $v$ to $u_f$
			\ForAll{$(X, j, \fail)$ in popped cache}
				\State \Call{addFail}{$L_f$, $i$, $j$}
			\EndFor
		\EndIf
	\EndIf
\EndFunction
\State ~
\Function{rtn}{nonterminal $X$, int $j$, int $h$}
	\If{$(X, j, h)$ not in popped cache}
		\State add $(X, j, h)$ to popped cache
		\ForAll{children $(L, i)$ of $(X, j)$ in the CRF}
			\If{$h \neq \fail$}
				\State \Call{addMatch}{$L$, $i$, $j$, $h$}
			\Else
				\State \Call{addFail}{$L$, $i$, $j$}
			\EndIf
		\EndFor
	\EndIf
\EndFunction
\end{algorithmic}
\end{figure}

In addition to call-stack management, a PEGLL parser-generator depends on helper functions to fill the descriptor queue $R$, descriptor cache $U$, and the BSR set $T$ representing the parse forest. 
The descriptor queue management functions are all in Figure~\ref{add-match-fail}.
\textsc{addMatch} and \textsc{addFail} enqueue the next descriptor after a nonterminal match or failure, accounting for the fact that lookahead expressions consume no input and adding a BSR element for matches. 
\textsc{addNt} is a convenience function which enqueues the first alternate of a nonterminal in the descriptor queue. 
\textsc{addDesc} checks if a given descriptor has already been processed, adding it to the descriptor queue and the cache of processed descriptors if not.

\begin{figure}
\caption{Code for descriptor queue management functions} \label{add-match-fail}
\begin{algorithmic}
\Function{addMatch}{slot $L$, int $i$, int $j$, int $h$}
	\State $T \gets T \cup {(L, i, j, h)}$
	\If{there is a lookahead expression before the dot in $L$}
		\State \Call{addDesc}{$L$, $i$, $j$}
	\Else
		\State \Call{addDesc}{$L$, $i$, $h$}
	\EndIf
\EndFunction
\State ~
\Function{addFail}{slot $L$, int $i$, int $j$}
	\If{there is a lookahead expression before the dot in $L$}
		\State \Call{addDesc}{$L$, $i$, $j$}
	\Else
		\State \Call{addDesc}{$L$, $i$, $i$}
	\EndIf
\EndFunction
\State ~
\Function{addNt}{nonterminal $X$, int $i$}
	\State $L \gets$ initial slot of first alternate of $X$
	\State \Call{addDesc}{$L$, $i$, $i$}
\EndFunction
\State ~
\Function{addDesc}{slot $L$, int $i$, int $h$}
	\If{$(L,i,h) \not\in U$}
		\State $R \gets R \cup {(L,i,h)}$
		\State $U \gets U \cup {(L,i,h)}$
	\EndIf
\EndFunction
\end{algorithmic}
\end{figure}

\subsection{Lexing}
Traditional maximal-munch lexers interact poorly with PEG parsers; in particular, the recursive-descent structure of a parsing expression grammar may impose context-sensitive priorities between tokens, and the usual scannerless design of a PEG may result in difficult-to-decompose token sets. 
Nonetheless, experience \cite{Ack19,Lau19} has shown that a separate lexing pass is a useful performance optimization.
As such, the PEGLL parser-generator uses a regular-expression lexer modified from the traditional maximal-munch approach \cite{Aetc07}. 
Rather than returning a single maximal-munch token at a given input position, as in the classical algorithm , the PEGLL lexer returns (from the \textsc{tokens}$(i)$ function) a map of all the tokens which match starting at position $i$ to their greatest right extent, defering the choice of which token to actually match to the PEG rules in the parser. 
The \textsc{testSelect} function, described in Figure~\ref{test-select} in the parser is used to check if any tokens in the FIRST set of the expression \cite{Red09} are also present in the token set returned from the lexer.

\begin{figure}
\caption[testSelect]{Code for \textsc{testSelect} function} \label{test-select}
\begin{algorithmic}
\Function{testSelect}{slot $L$, token set $T$, int $c_I$}
	\State $b \gets -1$
	\If{$L$ is nullable} $b \gets c_I$ \EndIf
	\ForAll{$(t,r)$ in $T$}
		\If{$r > b$ \textbf{and} $t \in FIRST(L)$} $b \gets r$ \EndIf
	\EndFor
	\State \Return{$b$}
\EndFunction
\end{algorithmic}
\end{figure}

\subsection{Syntactic Sugar} \label{syntax-sec}

The preceding discussion has outlined how PEG semantics are supported within the framework of the GLR algorithm; however, support for the fundamental PEG operators does not cover how several common PEG idioms are supported. 
Much like the EBNF notation for CFGs, PEGs typically support the ``syntactic sugar'' operators defined in Figure~\ref{sugar-fig} for optional and repeated matches. 
The $\alpha?$, $\alpha^*$, and $\alpha^+$ operators have their usual semantics of match $\alpha$ zero-or-one, zero-or-more, and one-or-more times, respectively, though note that the ordered-choice semantics of PEGs imply greedy match.
The PEGLL parser supports expressions using these operators by replacing such expressions during AST generation with a fresh nonterminal defined as in Figure~\ref{sugar-fig}.

\begin{figure}
	\centering
	\begin{equation*}
	\begin{aligned}[c]
	\Rule_{\alpha?} & = \alpha / \varepsilon \\
	\Rule_{\alpha^*} & = \alpha \Rule_{\alpha^*} / \varepsilon \\
    \Rule_{\alpha^+} & = \alpha \Rule_{\alpha^*} \\
	\end{aligned}
	\end{equation*}
	\caption[Syntactic sugar definitions]{Each syntactic sugar expression $\varphi$ is replaced by a fresh nonterminal $\Rule_\varphi$ defined as in this table.} \label{sugar-fig}
\end{figure}

\subsection{Unordered Choice}

PEGLL also supports an unordered-choice operator; the main challenge in implementing unordered choice in PEGLL is determining when to trigger the failure path for a nonterminal.
PEGLL introduces the failure paths in the FailCRF data structure to support the sequential nature of ordered choice -- when one alternate fails, it enqueues the next, and when the final alternate fails the algorithm marks the failure of the entire nonterminal. 
Unordered choice, by contrast, has concurrent semantics; all alternates can conceptually be executed in parallel (as in Scott \& Johnstone's GLL \cite{SJ10,SJ16}), and the nonterminal only fails if \emph{all} alternates fail. 

The approach taken by PEGLL is formalized in Figure~\ref{unordered-choice-algo}, but the essential idea is to duplicate all alternates of an unordered choice into a \emph{pass} alternate and a \emph{fail} alternate. 
Matching of an unordered choice begins with the first fail alternate, and proceeds sequentially through the alternates in source order, as in PEGLL's handling of ordered choice. 
Fail alternates all enqueue the next fail alternate on failure, or the next pass alternate on success, while pass alternates enqueue the next pass alternate on either success or failure; the final fail alternate marks the failure of the nonterminal on failure, while any alternate that succeeds marks the success of the nonterminal.
The effect of this duplication of alternates is to encode the failure or success of previous unordered alternates in the control-flow of the algorithm, rather than modifying the FailCRF to store this information in the data structure. 
This choice was made to avoid a memory-usage penalty in all FailCRF nodes to support unordered choice, but empirical investigation of the relative trade-offs of both approaches would be a fruitful direction for future work.

\begin{figure}
	\caption[Unordered choice code pattern]{Pattern for an unordered-choice nonterminal $X ::= \tau_1 | \ldots | \tau_p, p \geq 1$} \label{unordered-choice-algo}
	\begin{algorithmic}
	\Case{$X_{fail} ::= \cdot \tau_1$}
		\State $\langle$ generate code for $X ::= \cdot \tau_1$ with failure path $X_{fail} ::= \cdot \tau_2$ $\rangle$
		\State \Call{rtn}{$X$, $c_U$, $c_I$}; 
		\State $(L, c_I, t) \gets (X_{pass} ::= \cdot \tau_2, c_U, t')$ \Goto{nextSlot}
	\EndCase
	\Case{$X_{pass} ::= \cdot \tau_2$}
		\State $\langle$ generate code for $X ::= \cdot \tau_2$ with failure path $X_{pass} ::= \cdot \tau_3$ $\rangle$
		\State \Call{rtn}{$X$, $c_U$, $c_I$}; 
		\State $(L, c_I, t) \gets (X_{pass} ::= \cdot \tau_3, c_U, t')$ \Goto{nextSlot}
	\EndCase
	\State $\vdots$
	\State ~
	\Case{$X_{fail} ::= \cdot \tau_p$}
		\State $\langle$ generate code for $X ::= \cdot \tau_p$ with failure path $X_{fail} ::= \cdot \varnothing$ $\rangle$
		\State \Call{rtn}{$X$, $c_U$, $c_I$} \Goto{nextDesc}
	\EndCase
	\Case{$X_{fail} ::= \cdot \varnothing$}
		\State \Call{rtn}{$X$, $c_U$, $\fail$} \Goto{nextDesc}
	\EndCase
	\end{algorithmic}
	\end{figure}

\section{Related Work}
There are a number of pre-existing approaches to PEG parsing; Ford~\cite{For02} introduced the PEG formalism and two algorithms, a backtracking \emph{recursive descent} algorithm and a memoized \emph{packrat} algorithm. 
In practice, recursive decent is more efficient, but has exponential worst-case runtime, while packrat trades increased space and time in the common case for linear worst-case runtime\cite{Mos17}; most subsequent work has attempted to mitigate one or both of these shortcomings. 
Mizushima~\etal~\cite{MMY10} use \emph{cut operators} to reduce the memory usage of packrat parsing, while Kuramitsu~\cite{Kur15} and Redziejowski~\cite{Red07} use heuristic table-trimming mechanisms to similar effect. 
Medeiros \& Ierusalimschy~\cite{MI08} and Henglein \& Rasmussen~\cite{HR17} have developed a parsing machine approach and a tabular parsing algorithm, respectively, both of which have some evidence of efficient performance. 
Moss~\cite{Mos20} and Garnock-Jones~\etal~\cite{GJWE18} have independently developed derivative parsing algorithms for PEGs, neither of which claim improved performance, but which allow use of PEGs with algorithmic tools based on derivative parsing. 
Similarly to this work, Chida \& Kuramitsu \cite{CK17} have extended the PEG formalism to include unordered choice.
Medeiros~\etal \cite{MMI14left} and Hutchinson \cite{Hut20} have developed PEG parsers that accept left-recursive grammars, while Redziejowski \cite{Red21} has developed a grammar re-writing scheme that accomplishes the same end.

Scott \& Johnstone's GLL algorithm \cite{SJ10,SJ16} is based on the GLR algorithm of Tomita\cite{Tom85}. 
The PEGLL implementation (based on Ackerman's GoGLL \cite{Ack19}) also uses a modified call-return forest (CRF) for control flow and a binary subtree representation (BSR) for parse results, both as in Scott~\etal \cite{SJvB19}. 
Afroozeh \& Izmaylova \cite{AI15} have contributed performance improvements and extensions to the original GLL algorithm.
Parr~\etal \cite{PHF14} have developed the \emph{ALL(*)} parsing algorithm, with similar power to GLL but better reported runtime performance.

\section{Conclusion \& Future Work}

This paper has described PEGLL, a new algorithm for parsing parsing expression grammars, based on Scott \& Johnstone's GLL \cite{SJ10,SJ16}. 
The primary contribution is the FailCRF data structure, which adds failure paths to GLL's call-return forest, and code-generation algorithms to implement PEG semantics on top of this data structure. 
In addition to supporting the fundamental ordered-choice and lookahead operators of PEG, PEGLL also includes support for repetition operators as syntactic sugar, and also for CFG-style unordered choice using a duplication approach to track whether any of the unordered alternates have matched. 

An Apache-licenced implementation of PEGLL is available on GitHub\footnote{\url{https://github.com/bruceiv/pegll}}; this implementation passes the research team's internal correctness tests, but performance testing is left to future work. 
In addition to measuring and optimizing the performance of the PEGLL algorithm, a straightforward piece of future work would be to expand the expressivity of PEGLL's operator set; parenthesized subexpressions would be a natural expansion, as would more variants on the repetition operators -- in particular, non-greedy variants could be implemented by using unordered choice instead of ordered choice in the generated nonterminals. 

A more ambitious advance would be to support left-recursive rules in PEGLL. 
Combined with PEGLL's existing support for unordered choice, allowing left-recursive rules would make the set of languages parsed by PEGLL a superset of both parsing-expression languages and context-free languages, allowing language designers to apply the idioms and capabilities of both formalisms as desired. 
Given that GLL already supports left-recursive rules \cite{SJ10} and a number of approaches have been suggested for PEG-parsing as well \cite{MMI14left,Hut20,Red21}, this goal should be achievable.

\bibliographystyle{ieeetr}
\bibliography{deriv_parsing}

\end{document}